\documentstyle[preprint,aps]{revtex}

\begin{document}
\draft
\tighten

\title{The Feynman Variational Principle in the Worldline Representation
of Field Theory}

\author{R.~Rosenfelder $^1$, C.~Alexandrou $^{2}$
and A.~W.~Schreiber $^{3}$}

\address{$^1$ Paul Scherrer Institute, CH-5232 Villigen PSI, Switzerland\\
$^2$ Department of Natural Sciences, University of Cyprus,
CY-1678 Nicosia, Cyprus\\
$^3$ Department of Physics and Mathematical Physics and
           Research Centre for the Subatomic Structure of Matter,
           University of Adelaide, Adelaide, S. A. 5005, Australia }

\maketitle

\begin{abstract}
Following Feynman's treatment of the non-relativistic polaron problem, 
similar techniques are used to study relativistic field theories: after 
integrating out the bosonic degrees of freedom the resulting effective 
action is formulated in terms of particle trajectories (worldlines) 
instead of field operators.
The Green functions of the theory are then approximated variationally on
the pole of the external particles by using a retarded quadratic trial 
action. Application to a scalar theory gives non-perturbative, covariant
results for vertex functions and scattering processes. Recent progress in 
dealing with the spin degrees of freedom in fermionic systems, in particular
Quantum Electrodynamics, is discussed. We evaluate the averages needed 
in the Feynman variational principle for a general quadratic trial action 
and study the structure of the dressed fermion propagator.
\end{abstract}

\section{Introduction: The Polaron Problem}
Variational principles are widely used in classical mechanics, atomic,
molecular and nuclear physics but rarely in relativistic field theory.
However, there exists a successful example in a non-relativistic field 
theory, the so-called polaron problem. This describes an electron which 
slowly moves through an ionic crystal and polarizes its surroundings: the 
lattice vibrations (phonons) act back on the electron and change its 
properties. Feynman \cite{Fey} realized that the phonons can be integrated 
out exactly in the path integral giving rise to an effective
two-time action for the electron only. He then applied a variational
principle based on Jensen's inequality
\begin{equation}
\int {\cal D}x  \, e^{-S} 
 \>  \geq  \>  \int {\cal D} x \, e^{-S_t} \cdot
\exp \Bigl (-  \left < S - S_t \right >_{S_t} \Bigr )  
\label{Jensen}
\end{equation}
where $ \left <  \ldots \right >_{S_t} $ denotes an average  with respect 
to a suitable
trial action $S_t$. Feynman choose a {\it retarded quadratic} action and 
obtained excellent results for the ground state energy. Actually
his is the best analytic method which works for {\it all}
coupling constants: for weak coupling one recovers the 1-loop
result whereas at large coupling the maximal deviation is only  
2.2~\% \cite{polaron}.

\section{Scalar Theory: The Relativistic Polaron}
We have extended Feynman's polaron approach to a simple relativistic theory, 
the scalar Wick-Cutkosky model which describes heavy particles
(``nucleons'') interacting by the exchange of light particles 
(``mesons'') via a Yukawa coupling 
$ \> {\cal L}_{\rm int} \> = \> - g \, \Phi^2 \, \varphi \> $.
Starting from the generating functional for the Green functions of the theory
the following steps lead to a ``relativistic polaron''

\begin{itemize}
\item[(a)] Integrate out the heavy particle $\Phi$ and neglect the 
determinant, i.e. pair production (``quenched approximation'').

\item[(b)] Use the proper time representation to exponentiate 
\begin{equation}
\hspace*{-0.2cm} \frac{1}{ \hat p^2 - M^2 - 2 g \varphi(\hat x) + i 0} =
 \int_0^{\infty} \! \! \frac{dT}{2 i\kappa_0 } \>
\exp \left \{ \frac{i T }{2 \kappa_0} \left [
\hat p^2 - M^2 - 2 g \varphi(\hat x) \right ] \right \}  
\label{prop time repr}
\end{equation}
where $\kappa_0 > 0$ reparametrizes the proper time without changing the 
physics. We note that $\kappa_0 \to i$ and $p^2 \to - p^2$ gives the 
corresponding expression in Euclidean space.

\item[(c)] Write the $x$-space matrix elements of Eq. (\ref{prop time repr})
as a path integral over the particle trajectory $ x_{\mu}(t) $ (``particle or
worldline representation'').

\item[(d)] Perform the integration over the meson field $\varphi$ to get an 
effective, retarded two-time action in terms of the nucleon trajectory only.

\item[(e)] Make a variational approximation using the Feynman-Jensen 
variational principle with a retarded, quadratic trial 
action $S_t$. For  $ T \to \infty$ the 2-point function $ G_2(p)$ 
develops a pole at $p^2 = M^2$ and one finds the following relation 
between bare mass $M_0$ and physical mass $M$
\begin{equation}
- M_0^2 \> = \>   (\lambda^2 - 2 \lambda) M^2 + 2 \Bigl \{ \> \Omega[A] + 
V_{\rm WC} \, [\mu^2] \> \Bigr \} \> .
\label{Mano eq}
\end{equation}
This we call {\it Mano's equation} because K. Mano was the first to study
the scalar field theory with polaron methods \cite{Mano}.
\end{itemize}

\noindent
In Eq. (\ref{Mano eq}) $\lambda$ is a variational parameter and 
$ \Omega[A] $ and $V_{\rm WC} [\mu^2]$ denote the kinetic and potential 
term, respectively. The first one arises from the quadratic fluctuations
while the latter one is obtained by averaging the interaction term with the 
trial action. $ A(E)$ denotes the variational ``profile function'' and 
$ \mu^2(\sigma) $ the ``pseudotime'' which is related to $A(E)$ via a 
nonlinear integral transformation. For the explicit expressions and 
for more details see Refs. \cite{WC}. 
The Minkowski space version of
Eq. (\ref{Jensen}) only guarantees stationarity but one can 
show that the r.h.s. of Eq. (\ref{Mano eq}) is also minimal. In either case
one may now vary Mano's 
equation with respect to $\lambda$ and $A(E)$ to find 
the optimal values. This has been
done numerically for the 2-point function and extended to the case of 
an arbitrary number of mesons in the initial/final state so that form factors,
meson-nucleon scattering and deep inelastic scattering from the nucleon
could be calculated in a non-perturbative manner~\cite{WC}.

\section{Spinor Theory: QED}
An application to a more realistic theory like Quantum Electrodynamics (QED)
requires the description of the spin degrees of freedom of the electron
in the path integral. It is well known that this can be done by using
anticommuting Grassmann variables and we have developed 
a particular 4-dimensional formalism for massive Dirac particles \cite{QED1}.
We then followed essentially the same steps as in the scalar case with a 
few modifications. For example, Eq. (\ref{prop time repr}) is replaced by the 
(super) proper time representation
\begin{equation}
\frac{1}{\hat {\Pi \hspace{-7pt}/} - M + i 0} = 
\int_0^{\infty} \! \! \! dT \! \int \! d\chi \,
\exp \left [  \, - \frac{i}{2 \kappa_0} ( M^2 T + M \chi) \, \right ]
\, \exp \left [  \, \frac{i}{2 \kappa_0} ( \hat {\Pi \hspace{-7pt}/}^2 T + 
\hat {\Pi \hspace{-7pt}/} \, \chi) \, \right ]
\end{equation}
($ \hat \Pi_{\mu} = \hat p_{\mu} - e A_{\mu} (\hat x) $) where we use the 
Berezin integrals $\int d\chi  =  0 \> , \> \int d\chi \> \chi  =  1 $. Near 
the pole the electron propagator takes the form
\begin{equation}
G_2(p) \> \longrightarrow \>   Z \, \frac{ p \hspace{-5pt}/ + 
M_{\rm up}}{p^2 - M^2 + i 0}
\label{prop near pole}
\end{equation}
where $ M_{\rm up}$ may be different from $M$
(see below) and Mano's equation reads
\begin{equation}
M^2 ( 2 \lambda - \lambda^2) - M_0^2 \> = \>  2 \Bigl \{ \> \Omega[A]
- \Omega[A_f]  + V_{\rm QED}\, [\mu_f^2,\mu^2] \> \Bigr \} \> .
\label{Mano down}
\end{equation}
Here $A(E)$ is the bosonic and $A_f(E)$ the fermionic profile function.
Using methods described in Ref. \cite{QEDproc} we have calculated recently
the average of the QED interaction term with respect to a general
trial action which is quadratic in orbital and spin variables. 
The result is
\begin{eqnarray}
V_{\rm QED} \! &=& \!  - 4 \pi \alpha \, \frac{\nu^{2 \epsilon}}{\kappa_0} \,
\int_0^{\infty} \! \! d\sigma \int \! \frac{d^Dk}{(2 \pi)^D} \,
\frac{1}{k^2 + i0}
\Biggl \{    \frac{D-1}{4} k^2 \left [ \, \left (
\dot \mu^2_f(\sigma) \right)^2 - \left ( \dot \mu^2(\sigma) \right)^2 \,
\right ] \nonumber \\
&& \hspace{0.5cm} + \lambda^2 M^2
+ \lambda^2 (D-2) \frac{ (k \cdot p)^2}{k^2} \, \Biggr \} \,
\exp \left \{ \frac{i}{2 \kappa_0} \left [ k^2 \mu^2(\sigma)+ 2 \lambda 
p \cdot k \right ] \, \right \}
\label{V QED}
\end{eqnarray}
where $ \alpha = e^2/(4 \pi)$ is the fine structure constant and
 $ \nu$ an arbitrary mass parameter needed in $ D = 4 - 2 \epsilon $ 
dimensions.
Examination of Eq. (\ref{V QED}) reveals that it is 
more singular at small proper times than $V_{\rm WC}$ in the 
(super-renormalizable) Wick-Cutkosky model making the renormalization 
procedure more difficult.

\noindent
The mass appearing in the numerator of Eq. (\ref{prop near pole})
is determined by
\begin{equation}
M_{\rm up} \> = \>  M_0 \, \left [ \, \lambda + \lambda' -
\lambda \lambda' \, + \lambda' \, V_{\chi} \, \right ]^{-1} 
\label{Mano up}
\end{equation}
where $\lambda'$ is a spin variational parameter analogous to $\lambda$.
Varying with respect to this parameter gives 
$ \> \lambda = 1 + V_{\chi} \> \Rightarrow \> M_{\rm up} = 
M_0/(1 + V_{\chi}) = M_0/\lambda $ 
if inserted into Eq. (\ref{Mano up}). We also have evaluated the spin part
of the interaction average and -- for the special case $A_f(E) = A(E)$ -- 
found that $ \> V_{\chi} =  - V_{\rm QED}/(\lambda M^2) \> . $ 
This is a manifestation of the exact {\it supersymmetry} between 
bosonic and fermionic variables in the worldline description of a spinning
particle \cite{QED1}. The kinetic terms $\Omega$ then cancel and one finds 
from Eq. (\ref{Mano down}) that
$\>  M = M_0/\lambda \> \Rightarrow \> M_{\rm up} = M \> . $ 
Thus supersymmetry guarantees that there is only {\it one} pole 
at $ p \hspace{-5pt}/ = M $ in
the dressed electron propagator as it should be. We are presently deriving 
and analyzing the corresponding variational equations.

\section{Summary}

The worldline representation of field theory leads to a big reduction in 
the number of degrees of freedom , e.g. $ \Phi(x_{\mu}), \varphi(x_{\mu})
\to x_{\mu}(t) $. This is essential for a successful application of the
Feynman variational principle which (for technical reasons) is 
restricted to quadratic, albeit retarded trial actions. In this way 
encouraging nonperturbative results have been obtained for a simple scalar 
field theory. Describing the spin degrees of freedom 
by Grassmann variables allows the extension to fermionic theories. 
Applications to problems like chiral symmetry breaking
in strong coupling QED or the anomalous magnetic moment of the electron 
may be envisioned but still a number of open problems have to be addressed.
Among these the consistent renormalization to all orders in our variational
scheme seems to be the most urgent one.


\begin{thebibliography}{99}

\bibitem{Fey} R. P. Feynman, Phys. Rev. {\bf 97}, 660 (1955).

\bibitem{polaron} C. Alexandrou and R. Rosenfelder, Phys. Rep. {\bf 215}, 1
(1992); Yang Lu and R. Rosenfelder, Phys. Rev. {\bf B 46}, 5211 (1992) .

\bibitem{Mano} K. Mano, Progr. Theor. Phys. {\bf 14}, 435 (1955).

\bibitem{WC} R. Rosenfelder and A. W. Schreiber, Phys. Rev. {\bf D 53}, 3337,
3354 (1996); Nucl. Phys. {\bf A 601}, 397 (1996);
A. W. Schreiber, R. Rosenfelder and C. Alexandrou, Int. J. Mod. 
Phys. {\bf E 5}, 681 (1996); 
C. Alexandrou, R. Rosenfelder and A. W. Schreiber,
Nucl. Phys. {\bf A 628}, 427 (1998); N. Fettes and R. Rosenfelder, 
Few-Body Syst. {\bf 24}, 1 (1998).

\bibitem{QED1} C. Alexandrou, R. Rosenfelder and A. W. Schreiber, 
hep-th/9809101 .

\bibitem{QEDproc} R. Rosenfelder, C. Alexandrou and 
A. W. Schreiber, PSI-PR-98-13 .
 

\end{thebibliography}
\end{document}